# Apparent Retrocausation As A Consequence of Orthodox Quantum Mechanics Refined To Accommodate The Principle Of Sufficient Reason

Henry P. Stapp

*Lawrence Berkeley National Laboratory*
*University of California, Berkeley, California 94720*

**Abstract**. The principle of sufficient reason asserts that anything that happens does so for a reason: no definite state of affairs can come into being unless there is a sufficient reason why that particular thing should happen. This principle is usually attributed to Leibniz, although the first recorded Western philosopher to use it was Anaximander of Miletus. The demand that nature be rational, in the sense that it be compatible with the principle of sufficient reason, conflicts with a basic feature of contemporary orthodox physical theory, namely the notion that nature's response to the probing action of an observer is determined by pure chance, and hence on the basis of absolutely no reason at all. This appeal to pure chance can be deemed to have no rational fundamental place in reason-based Western science. It is argued here, on the basis of the other basic principles of quantum physics, that in a world that conforms to the principle of sufficient reason, the usual quantum statistical rules will naturally emerge at the pragmatic level, in cases where the reason behind Nature's choice of response is unknown, but that the usual statistics can become biased in an empirically manifest and apparently retrocausal way when the reason for the choice is empirically identifiable. It is shown here that if the statistical laws of quantum mechanics were to be biased in a certain plausible way then the basically forward-in-time unfolding of empirical reality described by orthodox quantum mechanics would generate the appearance of backward-time-causation of the kind that have been reported in the scientific literature.

## INTRODUCTION

An article recently published by the Cornell psychologist Daryl J. Bem [1] in a distinguished psychology journal has provoked a heated discussion in the New York Times [2]. Among the discussants was Douglas Hofstadter who wrote that: "If any of his claims were true, then all of the bases underlying contemporary science would be toppled, and we would have to rethink everything about the nature of the universe."

It is, I believe, an exaggeration to say that if any of Bem's claims were true then "all of the bases underlying contemporary science would be toppled" and that "we would have to rethink everything about the nature of the universe". In fact, all that is required is a relatively small change in the rules, and one that seems reasonable and natural in its own right. The major part of the required rethinking was done already by the founders of quantum mechanics, and cast in more rigorous form by John von Neumann [3], more than seventy years ago.

According to the precepts of *classical* mechanics, once the physically described universe is created, it evolves in a deterministic manner that is completely fixed by mathematical laws that depend only on the present values of evolving physically described properties. There are no inputs to the dynamics that go beyond what is specified by those physically described properties. [Here *physically described properties* are properties that are specified by assigning mathematical properties to space-time points, or to very tiny regions.] The increasing knowledge of human beings and other biological agents enters only as an *output* of the physically described evolution of the universe, and even nature itself is not allowed to interfere with the algorithmically determined mechanistic evolution.

This one-way causation from the physical aspects of nature to the empirical/epistemological/mental aspects has always been puzzling: Why should "knowledge" exist at all if it cannot influence anything physical, and hence be of no use to the organisms that possess it. And how can something like an "idea", seemingly so different from physical matter, as matter is conceived of in classical mechanics, be created by, or simply *be*, the motion of physical matter?

The basic precepts of classical mechanics are now known to be fundamentally incorrect: they cannot be reconciled with a plenitude of empirical facts discovered and verified during the twentieth century. Thus there is no reason to demand, or believe, that those puzzling properties of the classically conceived world must carry over to the actual world, which conforms far better to the radically different precepts of quantum mechanics.

The founders of quantum theory conceived the theory to be a mathematical procedure for making practical predictions about future empirical-experiential findings on the basis of our present knowledge. According to this idea, quantum theory is basically about the evolution of knowledge. This profound shift is proclaimed by Heisenberg's assertion [4] that the quantum mathematics "represents no longer the behavior of the elementary particles but rather our knowledge of this behavior", and by Bohr's statement [5] that "Strictly speaking, the mathematical formalism of quantum mechanics merely offers rules of calculation for the deduction of expectation about observations obtained under conditions defined by classical physics concepts."

The essential need to bring "observations" into the theoretical structure arises from the fact that evolution via the Schrödinger equation, which is the quantum analog of the classical equations of motion, produces in general not a single evolving physical world that is compatible with human experience and observations, but rather a mathematical structure that corresponds to a smeared out mixture of increasingly many such worlds. Consequently, some additional process, beyond the one generated by Schrödinger equation, is needed to specify what the connection is between the physically described quantum state of the universe and empirical/experiential reality. Epistemological factors must thereby become connected to the mathematically described physical aspects of the quantum mechanical description of nature.

The founders of quantum mechanics achieved an important advance in our understanding of nature when they recognized that the mathematically/physically described universe that appears in our best physical theory represents not the world of material substance contemplated in the classical physics of Isaac Newton and his direct successors, but rather a world of *"potentia"*, or *"weighted possibilities"*, for our future acquisitions of *knowledge* [6]. It is not surprising that a scientific theory designed to allow us to predict correlations between our shared empirical findings should incorporate, as orthodox quantum mechanics does: 1), a natural place for *"our knowledge"*, which is both all that is really known to us, and also the empirical foundation upon which science is based; 2), an account of the process by means of which we acquire our *knowledge* of certain physically described aspects of nature; and 3), a statistical description, at the pragmatic level, of relationships between various features of the growing aspect of nature that constitutes "our knowledge". What is perhaps surprising is the ready acceptance by most western-oriented scientists and philosophers of the notion that the element of chance that enters quite reasonably into the *pragmatic* formulation of physical theory, in a *practical* context where many pertinent things may be unknown to us, stems from an occurrence of raw pure chance at the underlying *ontological* level. Ascribing such capriciousness to nature herself would seem to contradict the rationalist ideals of Western Science. From a strictly rational point of view, it not unreasonable to examine the mathematical impact of accepting, *at the basic ontological level*, Einstein's dictum that: "God does not play dice with the universe", and to attribute the effective entry of pure chance at the pragmatic level to our lack of knowledge of the *reasons* for the "choices on the part of nature" to be what they turn out to be.

These "random" quantum choices are key elements of orthodox quantum mechanics, and the origin of these choices is therefore a fundamental issue. Are they really purely random, as contemporary orthodox theory asserts? Or could they stem at the basic ontological level from sufficient reasons?

## IMPLEMENTING THE PRINCIPLE OF SUFFICIENT REASON

I make no judgment on the significance of the purported evidence for the existence of various retrocausal phenomena. That I leave to the collective eventual wisdom of the scientific community. I am concerned here rather with essentially logical and mathematical issues, as they relate to the apparent view of some commentators that scholarly articles reporting the existence of retrocausal phenomena should be banned from the scientific literature, essentially for the reason articulated in the New York Times by Douglas Hofstadter, namely that the actual existence of such phenomena is irreconcilable with what we now (think we) know about the structure of the universe; that the actual existence of such phenomena would require a wholesale abandonment of basic ideas of contemporary physics. That assessment is certainly not valid, as will be shown here. Only a limited, and intrinsically reasonable, modification of the existing orthodox quantum mechanics is needed in order to accommodate the reported data.

In order for science to be able to confront effectively purported phenomena that violate the prevailing basic theory what is needed is an alternative theory that retains the valid predictions of the currently prevailing theory, yet accommodates in a rationally coherent way the purported new phenomena.

If the example of the transition from classical physics to quantum physics can serve as an illustration, in that case we had a beautiful theory that had worked well for 200 years, but that was incompatible with the new data made available by advances in technology.

However, a new theory was devised that was closely connected to the old one, and that allowed us to recapture the old results in the appropriate special cases, where the effects of the nonzero value of Planck's constant could be ignored. The old formalism was by-and-large retained, but readjusted to accommodate the fact that pq-qp was non-zero. Yet there was also *a rejection of a basic classical presupposition*, namely the idea that a physical theory should properly be exclusively about connections between physically described material events. The founders of quantum theory insisted that their physical theory was a pragmatic theory -- i.e., was directed at predicting practically useful connections between empirical (i.e., experienced) events [7].

This original pragmatic Copenhagen QM was not suited to be an ontological theory, because of the movable boundary between the aspects of nature described in *classical* physical terms and those described in *quantum* physical terms. It is certainly not ontologically realistic to believe that the pointers on observed measuring devices are built out of *classically* conceivable electrons and atoms, etc. The measuring devices, and also the bodies and brains of human observers, must be understood to be built out of quantum mechanically described particles. That is what allows us to understand and describe many observed properties of these physically described systems, such as their rigidity and electrical conductance.

Von Neumann's analysis of the measurement problem allowed the quantum state of the universe to describe the entire physically described universe: everything that we naturally conceive to be built out of atomic constituents and the fields that they generate. This quantum state is described by assigning mathematical properties to spacetime points (or tiny regions). We have a deterministic law, the Schrödinger equation, that specifies the mindless, essentially mechanical, evolution of this quantum state. But this quantum mechanical law of motion generates *a huge continuous smear of worlds of the kind that we actually experience.* For example, as Einstein emphasized, the position of the pointer on a device that is supposed to tell us the *time* of the detection of a particle produced by the decay of a radioactive nucleus, evolves, under the control of the Schrödinger equation, into a *continuous smear of positions corresponding to all the different possible times of detection*; not to a single position, which is what we observe [8]. And the unrestricted validity of the Schrödinger equation would lead, as also emphasized by Einstein, to the conclusion that the moon, as it is represented in the theory, would be smeared out over the entire night sky, until the first observer of it, say a mouse, looked.

How do we understand this huge disparity between the representation of the universe evolving in accordance with the Schrödinger equation and the empirical reality that we experience?

A completely satisfactory physical theory must include a logically coherent explanation of how the mathematical/physical description is connected to the experienced empirical realities. This demands, in the final analysis, a theory of the mind-brain connection: a theory of how our idea-like knowing aspects are connected to our evolving physically described brains.

The micro-macro separation that enters into Copenhagen QM is actually a separation between what is described in quantum mechanical physical terms and what is described in terms of *our experiences* -- expressed in terms of our everyday concepts of the physical world, refined by the concepts of classical physics. ([9], Sec. 3.5.)

To pass from *quantum pragmatism* to *quantum ontology* one can treat all *physically described* aspects quantum mechanically, as Von Neumann did. He effectively transformed the Copenhagen pragmatic version of QM into a potentially ontological version by shifting the brains and bodies of the observers -- and all other physically described aspects of the theory -- into the part described in quantum mechanical language. The entire physically described universe is treated quantum mechanically, and *our knowledge*, and *the process by means of which we acquire our knowledge about the physically described world,* were elevated to essential features of the theory, not merely postponed, or ignored! Thus certain aspects of reality that had been treated superficially in the earlier classical theories -- namely "our knowledge" and "the process by means of which we acquire our knowledge" -- were now incorporated into the theory in a detailed way.

Specifically, each acquisition of knowledge was postulated to involve, first, an initiating probing action executed by an "observer", followed by "a choice on the part of nature" of a response to the agent's request (demand) for this particular piece of experientially specified information.

This response on the part of nature is asserted by orthodox quantum mechanics to be controlled by *random chance*, by *a throw of nature's dice*, with the associated probabilities specified purely in terms of physically described properties. These "random" responses create a sequence of collapses of the quantum state of the universe, with the universe created at each stage concordant with the new state of "our knowledge".

If Nature's choices conform strictly to these orthodox statistical rules then the retrocausal results reported by Bem cannot be accommodated. However, if nature is not capricious -- if God does *not* play dice with the universe -- but Nature's choices have sufficient reasons, then, given the central role of "our knowledge" in quantum mechanics, it becomes reasonable to consider the possibility that Nature's choices are not completely determined in the purely mechanical way specified by the orthodox rules, but can be *biased* away from the orthodox rules in ways that depend upon the character of the knowledge/experiences that these choices are creating. The results reported by Bem can then be explained in simple way, and nature is elevated from a basically physical process to a basically psycho-physical process.

The question is then: What sort of biasing will suffice? One possibly adequate answer is a biasing that favors positive experiences and disfavors negative experiences, where positive and negative refers to the "feel" of the experience: pleasing or displeasing.

In classical statistical physics such a biasing of the *statistics* would not produce the appearance of *retrocausation*. But in quantum mechanics it does! The way that the biasing of the forward-in-time quantum causal structure leads to *seemingly* "retrocausal" effects will now be explained.

## BACKWARD IN TIME EFFECTS IN QUANTUM MECHANICS

The idea that choices made now can influence what has already happened needs to be clarified, for this idea is, in some basic sense, incompatible with our idea of the meaning of time. Yet the empirical results of Wheeler's delayed-choice experiments [10], and the more elaborate delayed-choice experiments of Scully and colleagues [11] are saying that, *in some sense*, what we choose to investigate now can influence what happened in the past. This backward-in-time aspect of QM is neatly captured by an assertion made in the recent book "The Grand Design" by Hawking and Mlodinow: "We create history by our observations, history does not create us" [12].

How can one make rationally coherent sense out of this strange feature of QM?

I believe that the most satisfactory way is to introduce the concept of "process time". This is a "time" that is different from the "Einstein time" of classical deterministic physics. That classical time is the time that is joined to physically described space to give classical Einstein space-time. (For more details, see my chapter in "Physics and the Ultimate Significance of Time" SUNY, 1986, Ed. David Ray Griffiths. In this book three physicists, D. Bohm, I. Prigogine, and I, set forth some basic ideas pertaining to time. [13])

Orthodox quantum mechanics features the phenomena of collapses (or reductions) of the evolving quantum mechanical state. In orthodox Tomonaga-Schwinger relativistic quantum field theory [14,15,16], the quantum state collapses not on an advancing sequence of constant time surfaces (lying at a sequence of times $t(n)$, with $t(n+1)>t(n)$, as in nonrelativistic QM), but rather on an advancing sequence of *space-like surfaces Sigma(n)*. (For each *n*, every point on the spacelike surface *Sigma(n)* is spacelike displaced from every other point on *Sigma(n)*, and every point on *Sigma(n+1)* either coincides with a point on *Sigma(n)*, or lies in the open future light-cone of some points on *Sigma(n)*, but not in the open backward light-cone of any point of *Sigma(n)*.)

At each surface *Sigma(n)* a projection operator $P(n)$, or its complement $P'(n) = I-P(n)$, acts to reduce the quantum state to some part of its former self!

For each surface *Sigma(n)* there is an associated "block universe", which is defined by extending the quantum state on *Sigma(n)* both forward and backward in time via the unitary time evolution operator generated by the Schrödinger equation. Let the index n that labels the surfaces *Sigma(n)* be called "process time". Then for each instant *n* of process time a "new history" is defined by the backward-in-time evolution from the newly created state on *Sigma(n)*.

This new "effective past" is the past that smoothly evolves *into the future the quantum state (of the universe) that incorporates the effects of the psycho-physical event that just occurred*. As far as current predictions about the future are concerned it is *as if* the past were the "effective past": the former *actual* past is no longer pertinent because it fails to incorporate the effects of the psycho-physical event that just occurred.

In orthodox QM each instant of process time corresponds to an "observation": the collapse at process time *n* reduces the former quantum state to the part of itself that is compatible with the increased knowledge generated by the new observation. This sequential creation of a sequence of new "effective pasts" is perhaps the strangest feature of orthodox quantum mechanics, and the origin of its other strange features.

The *actual* evolving physical universe is generated by the always-forward-moving creative process. It is forward-moving in the sense that the sequence of surfaces *Sigma(n)* advances into the future, and at each instant *n* of process time some definite, never-to-be-changed, psycho-physical events happens. But this forward-moving creative process generates in its wake an associated sequence of effective pasts, one for each process time

*n. The conditions that define the effective past associated with process time n change the preceding effective past imposing a "final" condition that represents what happened at process time n.* It is this "effective past" that evolves directly into the future, and is the past that, from a future perspective, has smoothly evolved into what exists "now". The actual past is not relevant to a history of the universe that starts from now and looks back, and projects smoothly into the immediate future.

The "histories" approach to quantum physics focuses attention on histories, rather than the generation of the profusion of incompatible possibilities. Both the effective past and the history associated with process time *n* depend upon which experiment is performed at time *n* , and in quantum mechanics that choice of which experiment is performed at process time *n* is not determined by the quantum state at process time *n*: it depends upon the agent's "free choice" of which probing action to initiate, where the word "free" specifies precisely the fact that this choice on the part of the agent is not determined by the known laws of nature.

Two key features of von Neumann's rules are mathematical formalizations of two basic features of the earlier pragmatic Copenhagen interpretation of Bohr, Heisenberg, Pauli, and Dirac. Associated with each observation there is an initial "choice on the part of the observer" of what aspect of nature will be probed. This choice is linked to an empirically recognizable possible outcome "Yes", and an associated projection operator *P(n)* that, if it acts on the prior quantum state ρ, reduces that prior state to the part of itself compatible with the knowledge gleaned from the experiencing of the specified outcome "Yes".

The process that generates the observer's choice of the probing action is not specified by contemporary quantum mechanics: this choice is, *in this very specific sense*, a "free choice on the part of the experimenter." Once this choice of probing action is made and executed, then, in Dirac's words, there is "a choice on the part of *nature*": nature randomly selects the outcome, "Yes" or "No" in accordance with the statistical rule specified by quantum theory. If Nature's choice is "Yes" then *P(n)* acts on the prior quantum state ρ, and if nature's answer is "No" then the complementary projection operator *P'(n) = I-P(n)* acts on the prior state. Multiple-choice observations are accommodated by decomposing the possibility "No" into sub-possibilities "Yes" and "No".

## MATHEMATICAL DETAILS

The description of orthodox quantum mechanics given above is a didactic equation-free account of what follows from the equations of quantum measurement theory. Some basic mathematical details are given in this section.

The mathematical representation of the dynamical process of measurement is expressed by the two basic formulas of quantum measurement theory:

$$\rho(n+1)_Y = \frac{P(n+1)\rho(n)P(n+1)}{Tr(P(n+1)\rho(n)P(n+1))} \quad ,$$

and

$$< P(n+1) >_Y = Tr(P(n+1)\rho(n)P(n+1)) = Tr(P(n+1)\rho(n)).$$

Here the integer *"n"* identifies an element in the global sequence of probing "measurement" actions. The symbol $\rho(n)$ represents the quantum state (density matrix) of the observed physical system (ultimately the entire physically described universe, here assumed closed) immediately *after* the *nth* measurement action; *P(n)* is the (projection) operator associated with answer *"Yes"* to the question posed by the *nth* measurement action, and *P'(n) = I–P(n)* is analogous projection operator associated in the same way with the answer *"No"* to that question, with *"I"* the unit matrix. The formulas have been reduced to their essences by ignoring the unitary evolution *between* measurements, which is governed by the Schrödinger equation.

The expectation value *<P(n+1)>$_Y$* is the normal orthodox probability that nature's response to the question associated with *P(n+1)* will be *"Yes"*, and hence that *ρ(n+1)* will be *ρ(n+1)$_Y$* . In the second equation I have used the defining property of projection operators, *PP=P*, and the general property of the trace operator: for any *X* and *Y*, *Tr(XY) = Tr(YX)*. (The trace operation *Tr* is defined by: *Tr(M)* = Sum of the diagonal elements of the matrix *M*).

Of course, one cannot know the density matrix ρ of the entire universe. The orthodox rules tell us to construct a "reduced" density matrix by taking a partial trace over the degrees of freedom about which we are ignorant, and renormalizing. This eliminates from the formulas the degrees of freedom about which we are ignorant.

The trace operation is the quantum counterpart of the classical integration over all of phase space. The classical operation is a summation that gives equal a priori weighting to equal volumes of phase space. That is the weighting that is invariant under canonical transformations, which express physical symmetries. The quantum counterparts of the canonical transformations are the unitary transformations, which leave the trace unchanged. Thus the orthodox trace rules are the rational way to give appropriate weights to properties about which have no knowledge, namely by assuming that properties related by physical symmetries should be assigned equal a priori weights.

All this is just orthodox quantum mechanics, elaborated to give a rationally coherent ontological account compatible with the standard computational rules and predictions. [17].

But the assumption that nature gives equal weights to properties that we, in our current state of scientific development, assume should be given equal weights, does not mean that nature itself must give such properties equal weight. Two states of the brain that are assigned equal statistical weight by the orthodox trace rule may be very different in the sense that one corresponds to a meaningful experience and the other fails to corresponds to any meaningful experience. Classical mechanics postulates that experiential qualities can make no difference in the flow of physical events. But, since quantum mechanics places experiences in a much more central role than classical mechanics, there is no rationally compelling reason to postulate in quantum mechanics that nature, in the process of choosing outcomes of empirical questions posed by agents, must be oblivious to the experiential aspects of reality. That issue should be settled by empirical results, not by classical-physics-based prejudice.

Consider a situation in which: (1), an agent (the participant) observes a property that corresponds to a projection operator *P*; and (2), a dynamically independent random number generator (RNG) creates either the property represented by the projection operator *Q*, or the property represented by the complementary property *Q'=(I-Q)*. Suppose at some time after these properties have been created they are still confined to two different systems that have never interacted, so that *PQ=QP*, and $\rho = \rho(P)\,\rho(Q)$. Then the probability of getting the answer (*PYes*), given that (QYes) occurs, is:

*Trace PQρ/TraceQρ  =Tr P ρ(P)/Tr ρ(P),*

which is independent of *Q*: the probability of *P* does not depend on what the dynamically independent RNG does. If the two systems interact later, beginning at time *t*, then the propagation to a final later time *t' of getting the outcome PYes at the earlier time t* is obtained by the action of a unitary transformation: the quantity upon which the trace operation acts gets multiplied on the right by a unitary *U(t',t)*, and on the left by the Hermitian conjugate of *U(t',t)*. This action leaves the probability of PYes at the earier time t intact.

If there is then a final measurement at the later final time *t'* of a property represented by a projection operator *R*, then there could be a dependence upon whether Nature's choice at time *t'* actualizes *R* or *R' = (I-R)*. But if the orthodox random rules are obeyed then the net effect, obtained by averaging over the two properly weighted possibilities, is null – because *R +R' = I* : the mere fact that an observation is made at the final time *t'* has no effect on the correlation (actually the lack of correlation) between *P* and *Q* observed at the earlier time t. However, if Nature's choices are not weighted in the orthodox way then the contributions from the two "complementary" effective pasts, arising from *R* and *R'*, respectively, will be unequally weighted,  and the sum over the two terms might no longer wipe out the effects of the differing effective pasts that lead to Ryes and RNo, respectively: observable effects can arise from an excess of histories/probability corresponding to the option, say RYes, that Nature's choice favored, and a deficit of histories/probability corresponding to the option, say RNo, that Nature's choice disfavored. If Nature's choices are not weighted in accordance with the orthodox rules then the different histories leading to differing final states, RYes or RNo, can get biased weightings in the effective past actualized by Nature's biased choice between RYes and RNo.

## APPLICATION TO BEM'S EXPERIMENTS

All nine of Bem's experiments have the following general form: First, in each instance in a series of experimental instances, the participant is presented with some options, and picks a subset of these options as 'preferred'. These preferences are duly recorded. *Later*, for each instance, an emotional stimulus is applied to the participant. The stimulus, and the way it is applied to the participant, is determined by some random number generators (RNGs). These RNGs are, according to both classical and quantum ideas, dynamically independent of the participant's earlier actions. But Bem's empirical result is that the probability that an option is preferred by the participant at the early time depends upon choices made later by the RNGs.

This finding seems to suggest that either the believed-to-be dynamically independent RGNs are being influenced in a mysterious and complex way by the participant's earlier actions; or the participant's earlier actions are being affected in a complex retrocausal (backward-in-time causal) way by the choices made by RNGs.

The kinds of actions made by the participant, and by the RNGs, vary greatly over the nine experiments. But, from a quantum standpoint, one single presumption explains all of the reported results, and explains them all in a basically forward-in-time causal way, without any mysterious influence of the participant's choice of preference on the RGNs. This presumption is that the choices on the part of nature, which are essential elements of orthodox quantum mechanics, are slightly biased, relative to the orthodox quantum stastistical rules, in favor of the actualization of positive feelings in the mind of the participant, and against the actualization of negative feelings.

For example, in the first Bem experiment the participant is shown two similar screens, L and R, and is told that behind one screen lies a picture, and behind the other lies the image of a blank wall. S/he is instructed to choose a "preferred" screen, P (either L or R) behind which s/he *feels* the picture lies. *After* the participant's preference P, either L or R, is recorded, a first random number generator, RNG1, chooses a "target" screen T (either L or R), and assigns a *picture* to target screen T, and an image of a blank wall to the other screen. A second random number generator, RNG2, decides, with equal probabilities, whether the picture will be "Erotic" or "Neutral" (The stimulus type S is either E or N)). What has been determined by the RNGs to lie behind the preferred screen P is then shown to the participant.

Bem's empirical result is that the participants choose more often than orthodox quantum mechanics (or classical statistical mechanics) predicts the screen behind which *will lie* an erotic picture, but prefers L and R with equal probability if RNG2 chooses a "neutral" picture.

If the well-tested random number generators are working as they normally do then this empirical result would appear to be a case of retrocausation (causal action backward in time): the choices made *later* by the two RNGs are influencing the subject's *earlier* choice between L and R. An alternative possibility is that RGN2, which chooses between "erotic" and "neutral", is being influenced by the participant's earlier choice between L and R, so that the screen behind which the participant looks will tend to be erotic, but only if RNG1 chooses "picture" not "blank wall". These explanations would require, as Hofstadter remarked, a very major revision of contemporary ideas about how nature works.

These putative explanations, just described, are in terms of strange causal connections between the earlier choice made by the participant and the later choices made by the supposedly dynamically independent RNGs: the character of the participant's final feelings are not considered. But Bem's results are explained in natural, rational, and basically forward-causal way, without any weird behaviors of the RNGs, if *Nature's choice of the participant's final experience* – a choice that is an absolutely essential element of orthodox quantum theory, and the place where the quantum element of chance enters -- favors, relative to the statistical predictions of orthodox quantum mechanics, the occurrence of positive (pleasing) experiences and disfavors the occurrence of negative feelings. If such a biasing of Nature's choices were to occur, then the observed greater likelihood of the participant's choosing the screen, L or R, behind which an erotic picture *will lie* would arise from the enhanced likelihood that nature will actualize an erotic experience, instead of an experience of a neutral picture or a blank wall. In this experimental set up an erotic experience can occur only if P=T and S=E: the participant's earlier choice of the between L and R must agree with the later choice of RNG1 between L and R, since otherwise the participant will see only a blank wall, and even if P=T, the choice of stimulus S must be E, since otherwise the participant will see a neutral picture.

A compact way of stating this explanation is to say that the quantum histories [defined by the sequences of choices (P,T,S,F) leading to the final experience F=+, or F=-] that lead to F=+ are more likely to occur than the rules of orthodox quantum mechanics predict. Only those histories in which the two L/R choices agree (P=T) can lead to an Erotic experience, because if these two choices disagree the participant will see a blank wall. But this enhancement will occur only in the subset of histories in which S=E.

In Bem's Experiment 2, "Precognitive Avoidance of (Subliminal) Negative Stimuli", a sequence of similar pairs of neutral pictures is shown to the participant, who chooses a 'preferred' picture from each neutral pair. After each such recorded choice of preference P, a RNG1 makes a random choice of one picture from the initial pair. The picture chosen by RNG1 is called the 'target' T. Then the apparatus flashes a *subliminal* picture, the stimulus S, that is positive, S=+, if T=P, but is highly negative, S=-, if the preferred neutral picture P is *not* the subsequently randomly chosen target picture T.

The normal idea of forward causation does not allow this random choice of target, and the associated application of a stimulus, both of which occur *after* the recorded choice of preference, to affect, in any instance, the participant's previously recorded choice of preference between two matched neutral pictures. Yet Bem's predicted and empirically validated result is that the picture P preferred at an earlier time by a participant is more likely to be the subsequently chosen target picture T than the subsequently chosen nontarget, even though the choice between target and nontarget was 50-50 random, and was made only later. The non-targeted pictures, which are, according to Bem's empirical findings, less likely to be preferred than chance predicts, are the pictures that occur in conjunction with the later subliminal application to the participant of highly unpleasant pictures. Hence they should lead to unpleasant participant feelings and should therefore, according to the present hypothesis, be less likely than chance predicts to be selected by Nature's choice to become an actually experienced outcome:
<(P,T not P, S-, F-)> < <P, T=P. S+, F+>.

This experimental protocol is quite different from the protocol of the first experiment. In the first experiment the stimulus that was applied later to the participant was independent of the participant's earlier choice of preference, whereas in experiment 2 the stimulus that is applied later to the participant depends upon the earlier choice of preference. Moreover, the stimulus was supraliminal in the first experiment but subliminal in the second experiment.

Nevertheless, the apparently retrocausal effect in the second experiment follows from the same quantum assumption as before, namely that Nature's choice of which final experience actually occurs has a tendency to increase the likelihood of positive, and diminish the likelihood of negative, final experiences of the participant. In experiment 2 the effect of this biasing is to diminish the likelihood of instances in which the *final feeling* of the participant is negative, due to the earlier application to the participant of an (albeit subliminal) highly negative stimulus.

Bem's experiments 3 and 4 are "Retrocausal Primings". Unlike the first two experiments, they do not involve matched neutral pairs between which the participant must choose. Rather, each instance now involves a single picture, which is either positive or negative. This non-neutral picture is shown to the participant, who responds by pressing a first or second button according to whether s/he feels the picture to be pleasing or not. The *time* that it takes for the participant to react to the picture is recorded. *Then,* a 'word' is selected by a RNG, and is (supraliminally) shown to the participant. The previously recorded reaction time turns out to be shorter or longer according to whether feeling of the word is "congruent" or 'incongruent" to the feeling of the picture experienced earlier.

There is also a 'normal' version of the experiment in which the word chosen by the RNG is displayed *before* the participant chooses his preference. Bem's experimental set-up is one for which, also in the 'normal' version, the recorded reaction time is shorter or longer according to whether feeling of the word is "congruent" or 'incongruent" to the feeling of the picture shown earlier.

The question is: How, in the retrocausal version, can the *reaction time*, which was recorded *earlier,* depend on which word was randomly selected *later*?

This empirical finding is explained by an assumed biasing of "Nature's choice" of the participant's final feeling that favors congruency in the flow of experience over incongruency. Such a putative biasing of Nature's choice has the effect of adding to the effective past, after nature's biased choice, of an abnormal contribution that corresponds to the *addition* of extra histories that lead to the mentioned positive feelings, and to the subtraction of histories that lead to analogous negative feelings. These changes in the weightings of the differing histories, in accordance with the nature of the feeling induced by the stimulus word, have an effect on the quantum state of the participant's brain, during the process of his or her choosing between positive and negative pictures. This effect on the brain is roughly independent of whether the stimulus was applied before or after the participant's choice of response. In both cases the "effective past" state of the brain of the participant during his or her process of choosing a response is changed in essentially the same way: it is not important whether the change in the effective state of the participant's brain, during the process of choosing his or her preference, comes from changes in the earlier or later boundary condition on that "effective past" state of the brain. The key point is that, as discussed in earlier sections, the "effective past" incorporates the conditions imposed by the occurrence of the final outcome! A "history" starts from what is now known, and extends backward from the known present, which depends on nature's most recent choice.

In the three habituation experiments the participant is again shown a matched pair of pictures, and is asked which one s/he prefers. The two matched pictures are both strongly negative, both strongly positive (erotic), or both essentially neutral in the first, second, and third experiments, respectively. (I have slightly reorganized Bem's data in this way for logical clarity, and ignored some inconclusive data with small statistics in which the later stimuli were supraliminal.) After the participant makes a binary recorded choice of preference, an RNG chooses one of the two similar pictures as target, and the targeted picture is subliminally flashed several times. The subliminal re-exposures, made after the participant's choice of preference of the targeted emotion-generating picture, have the effect of reducing, in the case of the positive pairs of pictures, and increasing in the case of negative pairs of pictures, the fraction of instances in which the (previously) preferred picture was the target rather than the nontarget: the effective positivity/negativity of the targeted (and hence repeatedly subliminally represented) pictures was reduced. This is explained by a reduction in the emotional intensity of the participant's final feeling, caused by the repeated re-exposure to the highly emotional pictures, and the attendant diminuation of the biasing of Nature's choices.

In the two memory experiments the participant is exposed to a sequence of 48 common everyday nouns, and is then tested see which words s/he remembers. Afterwards, 24 of the original set of words are randomly chosen to be 'targets, and then, in a sequence of computer-controlled actions, the participant is repeatedly re-exposed to each of the target words, but none of the non-target words. It is subsequently found that among the recalled words there are more target words than non-target words. This is explained if Nature's choice of the participant's final feelings favors the feel of congruent streams of conscious experiences over the feel of less congruent ones.

All of Bem's reported results are thus explained by a single presumption, namely that Nature's choices, rather than being strictly random, in accordance with the rules of contemporary orthodox quantum mechanics, are slightly biased, relative to the predictions of orthodox quantum mechanics, in favor of outcomes that feel pleasing, and against outcomes that feel displeasing.

This explanation is "scientific", in the sense that it can be falsified. If the output of the RNGs were to be observed by an independent observer, *before* the RNG-chosen action is made on the participant, then the biasings reported by Bem should disappear, because Nature's choice would then be about the possible experiences of the independent observer rather than about those of the participant. A more elaborate test would be to have two participants doing the experiment on the same sequence of picture, with reversed polarities. A dependence upon who first experiences the output of the RNG would, if it were to occur, constitute spectacular support for the notion that our experiences really do influence the course of physically described events, rather than being merely causally inert by-products of a process completely determined by purely physical considerations alone.

In the above discussion I have treated all of the RNGs as true quantum-process-based random number generators. In some of the experiments the RNG was actually a pseudo-random number generator, a PRNG. In principle a PRNG is, in these experiments, just as good as a true RNG, unless *at the time of its effective action* some real observer *actually knows* everything needed to specify what the pseudo-random choice must be. Unless the outcome is actually specified by what is actually currently known by observing agents, the outcome is, within this orthodox framework, effectively undetermined.

# CONCLUSION

Numerous reported seemingly backward-in-time causal effects could be naturally explained within a slightly modified version of orthodox forward-in-time quantum mechanics. In this version, Nature's "random" choices of which outcomes to actualize are slightly biased in favor of actualizing positive feelings and against actualizing negative feelings. In the Bem experiments the power of a participant to bias the statistics depends very strongly on the participant's mental characteristics, as measured by simple questionnaire responses.

ACKNOWLEDGEMENTS

This work was supported by the Director, Office of Science, Office of High Energy and Nuclear Physics, of the U.S. Department of Energy under contract DE-AC02-05CH11231

# REFERENCES

[1] D. J. Bem, *Feeling the Future: Experimental Evidence for Anomalous Retroactive Influences on Cognition and Affect*. Journal of Personality and Social Psychology, **100**, 407-425. (2011).

[2] New York Times (http://www.nytimes.com/roomfordebate/2011/01/06/the-esp-study-when-science-goes-psychic/a-cutoff-for-craziness).

[3] J. von Neumann, *Mathematical Foundations of Quantum Theory*. (Princeton University Press, Princeton N.J., 1955).

[4] W. Heisenberg. The representation of Nature in contemporary physics. *Daedalus,* **87**, 95-108 (summer, 1958).

[5] N. Bohr, *Essays 1958/1962 on Atomic Physics and Human Knowledge*. (Wiley, New York, 1963) p.60.

[6] W. Heisenberg, in Appendix A of ref. [7].

[7] H. P. Stapp, The Copenhagen Interpretation. *American Journal of Physics*, **40**, 1098-1116 (1972). Reprinted as Chapter 3 in *Mind, Matter, and Quantum Mechanics.* (Springer-Verlag, Berlin, Heidelberg, New York, 1993, 2003, 2009).

[8] A. Einstein in *Albert Einstein:Philosopher-Scientist.* ed. P. A. Schilpp. (Tudor, New York, 1951) p. 670.

[9] N. Bohr, *Essays 1958/1962 on Atomic Physics and Human Knowledge*. (Wiley, New York, 1963) p. 60.

[10] J. A. Wheeler, The 'past' and the delayed-choice double-slit experiment. In: A. R. Marlow ed. *Mathematical Foundations of Quantum Theory.* (Academic Press, New York, 1987) p. 9-48.

[11] Y. H. Kim, R. Yu, S.P. Kulik, Y. Shih, and M.O. Scully, Delayed "choice" quantum eraser. *Phys. Rev. Lett.,* 84, 1-5, (2000).

[12] S. Hawking and L. Mlodinow, *The Grand Design*. (Bantam, New York, 2010) p.140.

[13] H.P. Stapp, Process Time and Einstein Time, in *Physics and the Ultimate Significance of Time*, ed. David Ray Griffiths. (SUNY Press, Albany. 1986)

[14] S. Tomonaga, On a relativistically invariant formulation of the quantum theory of wave fields. *Progress in Theoretical Physics*, 1, 27-42 (1946)

[15] J. Schwinger, The theory of quantized fields I. *Physical Review*, 82, 914-927 (1951)

[16] H. P. Stapp, *Mindful Universe: Quantum Mechanics and the Participating Observer.* (Springer-Verlag, Berlin, Heidelberg 2007, 2011). Chapter 13.

[17] H. P. Stapp, Quantum Theory and the Role of Mind in Nature. *Foundations of Physics*, 31, 1465-99 (2001) [arxiv.org/abs/quant-ph/0103043]

The answer is the same as before: the effective background state of the subject's brain-- positive or negative -- in the extra effective histories created by nature's *biased* response to the later priming is similar to the normal forward-in-time effect of the same priming: it matters little whether this influential background state of the brainis produced by an initial or a final boundary condition on the effective past.

To achieve this explanation one needs to relax the condition that the von Neumann process-1 action of posing a question identifies a property of the brain of the observing system that can be grasped as a high-grade conscious experience. One might replace "high-grade conscious experience" by "experienced mood", which could be generated by the subliminal stimuli. Or high-grade conscious experience could be replaced by a lower-level kind of experience. One actually needs such a relaxing anyway, in order to allow lowly life-forms to enter into the quantum dynamics.
The general proposal that Nature's choices arise from reasons should be helpful also to the effort to understand the origin of life. If nature exhibits a slight biasing for positive experiences of individual human beings it would perhaps be natural for it exhibit a large bias in favor of the existence in the universe of systems that can represent meaning, and act on the basis of such reasons. What is at issue here is the basic nature of the logical break, in the passage to a quantum universe, with the classical-physics conception of a mindless, purposeless, reasonless, purely mechanical universe. Experiments of the kind performed by Bem, and variations thereof, if they stand the test of time, have the potential of shedding important scientific light on this question.